\documentclass[journal]{IEEEtran}

\ifCLASSINFOpdf
\else
   \usepackage[dvips]{graphicx}
\fi
\usepackage{url}
\usepackage[nolist]{acronym}
\usepackage{graphicx}
\usepackage{tikz}
\usetikzlibrary{arrows.meta}

\tikzset{%
  >={Latex[width=2mm,length=2mm]},
            base/.style = {rectangle, rounded corners, draw=black,
                           minimum width=4cm, minimum height=1cm,
                           text centered, font=\sffamily},
  activityStarts/.style = {base, fill=blue!30},
       startstop/.style = {base, fill=red!30},
    activityRuns/.style = {base, fill=green!30},
         process/.style = {base, minimum width=2.5cm, fill=orange!15,
                           font=\sffamily},
}

\begin{document}

\title{Beyond the Labels: Unveiling Text-Dependency in~Paralinguistic Speech Recognition Datasets}

\author{Jan Pešán, Santosh Kesiraju, Lukáš Burget and Jan ''Honza'' Černocký
\thanks{Jan Pešán, Santosh Kesiraju, Lukáš Burget and Jan Černocký are with the Faculty of Information Technology, Brno University of Technology, Brno, 61200, Czechia (e-mails: ipesan@fit.vutbr.cz, kesiraju@fit.vutbr.cz, burget@fit.vutbr.cz, cernocky@fit.vutbr.cz).}
\thanks{The work was supported by European Union’s Horizon Europe project No. SEP-210943216 "ELOQUENCE", and Czech National Science Foundation (GACR) project "NEUREM3" No. 19-26934X. Computing on IT4I supercomputer was supported by the Ministry of Education, Youth and Sports of the Czech Republic through e-INFRA CZ (ID:90254).}
}

\markboth{Signal Processing Letters}{Shell \MakeLowercase{\textit{et al.}}: Bare Demo of IEEEtran.cls for IEEE Journals}
\maketitle

\begin{abstract}
Paralinguistic traits such as cognitive load and emotion are recognized as important areas in speech recognition research, often examined through specialized datasets such as CLSE and IEMOCAP. However, the integrity of these datasets is seldom scrutinized for text dependency. This paper critically evaluates the assumption that machine learning models trained on such datasets genuinely learn to identify paralinguistic traits, rather than merely capturing lexical features. By examining the lexical overlap in these datasets and testing the performance of machine learning models, we expose significant text dependency in trait labeling. Our results suggest that some machine learning models, especially large pre-trained models such as HuBERT, might inadvertently focus on lexical characteristics rather than the intended paralinguistic features.
\end{abstract}


\IEEEpeerreviewmaketitle

\section{Introduction}
\IEEEPARstart{W}{hile} the primary focus of speech recognition research gravitates towards \ac{ASR}, the study of paralinguistic traits, such as cognitive load, physiological stress, and emotions, remains a significant field too. These traits are of interest for applications ranging from human-computer interaction to psychological research and rely heavily on dedicated datasets. Two key datasets are \ac{CLSE}~\cite{yap2012speech-egg-cognitive-load}, commonly used for cognitive load recognition, and \ac{IEMOCAP}~\cite{Busso2008-iemocap}, primarily employed for emotion recognition.

A prevailing assumption is that machine learning algorithms trained on these datasets learn to recognize paralinguistic traits based on observable physiological or psychological changes in speech production. However, this paper challenges this assumption by revealing a critical oversight: we provide evidence of significant lexical correlation between the labels (e.g., cognitive load or emotion) and the uttered sentences within these datasets. 

We analyze \ac{CLSE} and review the design of \ac{IEMOCAP} to substantiate this claim. Given the implications of our findings, the paper serves as a call reevaluation of existing datasets and methodologies to ascertain that machine learning systems are learning to recognize what they are designed to recognize.

\section{Related Work}

\subsection{Datasets in Focus}
Two key datasets facilitate paralinguistic studies:
\begin{itemize}
    \item \textbf{CLSE (Cognitive Load with Speech and EGG)}: This is intended to test recognition of cognitive load by incorporating both speech recordings and \ac{EGG} signals.
    \item \textbf{IEMOCAP (Interactive Emotional Dyadic Motion Capture)}\footnote{\url{https://sail.usc.edu/iemocap/index.html}}: It covers about 12 hours of scripted and spontaneous dialogues. It captures speech and facial and hand movements to study a range of emotions.
\end{itemize}

\subsection{Machine Learning Approaches to Paralinguistics}
Various machine learning algorithms, from classic UBM-iVector~\cite{compare2014-ivector}, through LSTM~\cite{gallardo2019-saliency-based-lstm-on-clse} to recent large pre-trained models like wav2vec~\cite{wav2vec-clse}, have been employed on these datasets. The general presumption is that these models are capturing paralinguistic features rooted in physiological or psychological changes~\cite{Schuller2013-computational-paralinguistics,MaryZarate2015-linguistic-paralinguistic-features}. 

Pepino~et~al~\cite{pepino} discovered the text-dependency issue in their work on Emotion Detection on \ac{IEMOCAP}. In their experiments they observed over-optimistic results while using the original train-test splits. Yuanchao~et~al~\cite{emotions-corpora-study} analyzed wav2vec performance on \ac{IEMOCAP} and found that this model prioritizes linguistic content over para-linguistic information.

In paralinguistic research, it is essential to consider both the content and the manner of speech. Physiological and prosodic features often capture the manner of speech, while lexical choices, semantics, and syntax provide insights into the content. Ignoring either aspect may lead to an incomplete understanding of paralinguistic traits. This study emphasizes the importance of acknowledging lexical correlates when analyzing paralinguistic datasets to ensure comprehensive evaluation and accurate model training.

\section{Hypothesis and Methodology}

\subsection{Hypothesis}
Our hypothesis posits that machine learning models trained on paralinguistic speech recognition datasets like \ac{CLSE} and \ac{IEMOCAP} may exhibit significant text dependency. This dependency could undermine the models' ability to genuinely learn paralinguistic traits, instead focusing on lexical characteristics of the speech data.

\subsection{Methodology}
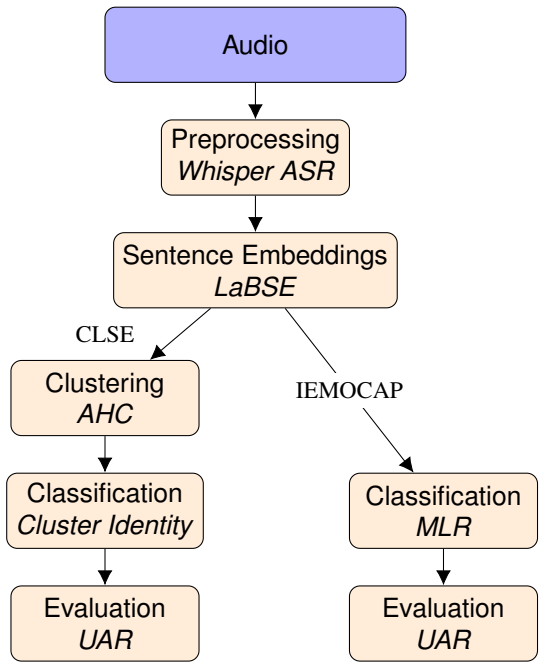
\begin{figure}
\centering
\begin{tikzpicture}[node distance=1.5cm,
      every node/.style={fill=white, font=\small}, align=center]
    \node (start)             [activityStarts]              {Audio};
    \node (preproc)           [process, below of=start]              {Preprocessing \\ \textit{Whisper ASR}};
    \node (sentenceEmbeddings)           [process, below of=preproc]              {Sentence Embeddings \\ \textit{LaBSE}};
    \node (CLSECLustering)    [process, below of=sentenceEmbeddings, xshift=-2cm, yshift=-0.2cm]          {Clustering \\ \textit{AHC}};
    \node (CLSEClassification) [process, below of=CLSECLustering]          {Classification \\ \textit{Cluster Identity}};
    \node (CLSEEvaluation) [process, below of=CLSEClassification]          {Evaluation \\ \textit{UAR}};

    \node (IEMOCAPClassification) [process, right of=CLSEClassification, xshift=3cm]           {Classification \\ \textit{MLR}};
    \node (IEMOCAPEvaluation) [process, right of=CLSEEvaluation, xshift=3cm]           {Evaluation \\ \textit{UAR}};

    \draw[->]             (start) -- (preproc);
    \draw[->]             (preproc) -- (sentenceEmbeddings);
    \draw[->]             (sentenceEmbeddings) -- node [xshift=-1cm]{CLSE} (CLSECLustering);
    \draw[->]             (CLSECLustering) -- (CLSEClassification);
    \draw[->]             (CLSEClassification) -- (CLSEEvaluation);
    \draw[->]             (IEMOCAPClassification) -- (IEMOCAPEvaluation);
    \draw[->]             (sentenceEmbeddings) -- node {IEMOCAP} (IEMOCAPClassification);
  \end{tikzpicture}
  \caption{Block scheme of the experiment structure for different datasets}
  \label{fig:scheme}
\end{figure}

To test this hypothesis, we conducted experiments on both the \ac{CLSE} and \ac{IEMOCAP} datasets. We employed various machine learning models, including classic and state-of-the-art approaches, and analyzed their performance with and without lexical overlap. Our methodology involves steps seen in Figure~\ref{fig:scheme}.

\section{Dataset and Task Description}

\subsection{CLSE}
Our primary study focuses on the main part of the \ac{CLSE} database, termed as CLSE-Span, which employs the~\ac{RSPAN}~\cite{Daneman1980-rspan}. In \ac{RSPAN}, participants are asked to validate sentences for their logical coherence while memorizing a sequence of letters. Sentences are randomly chosen from a closed set. The task is divided into sets, where each set contains varying trials of sentence validation and letter memorization. The database consists of 21 such sets with each participant producing 75 utterances, totaling 1800 utterances. The average utterance duration is four seconds. Labels for the cognitive load values are: low, medium, and high load.

\subsection{IEMOCAP}
The \ac{IEMOCAP} dataset comprises 151 dialog videos, featuring a pair of speakers in each session, resulting in a total of 302 individual videos with approximately 12 hours of data. Ten actors were recorded in dyadic sessions (five sessions with two subjects each). They were asked to perform three selected scripts with clear emotional content. In addition to the scripts, the subjects were also asked to improvise dialogs in hypothetical scenarios, designed to elicit specific emotions (happiness, anger, sadness, frustration, and neutral state). The original annotations were expanded by the authors after the collection. The new annotated classes contain nine distinct emotions (\textit{angry, excited, fear, sad, surprised, frustrated, happy, disappointed, and neutral}), along with metrics for valence, arousal, and dominance.

\section{Experimental Setup}

\subsubsection{Data Preprocessing}
We used Whisper ASR~\cite{radford2022robust-whisper}, a pre-trained multilingual \textit{large-v3} model, to transcribe the speech data from both datasets. Subsequently, we utilized Sentence Transformer embeddings as features, using the \ac{LaBSE} model~\cite{feng-etal-2022-language-labse}. LaBSE utilizes a dual-encoder architecture trained in two steps: first on 17 billion monolingual sentences via Masked Language Modeling, and then further refined on six billion translation pairs covering 109 languages. The final model is publicly accessible\footnote{\url{https://huggingface.co/sentence-transformers/LaBSE}} and was used to extract 768-dimensional sentence embeddings.

\subsubsection{Clustering}
For the \ac{CLSE} dataset, we applied \ac{AHC} to the sentence embeddings to group similar sentences. After manually correcting minor clustering errors, we identified 81 different sentences. Each sentence was assigned an ID based on its cluster membership. 

\subsubsection{Classification}
Using the clustered data, we assigned the most frequent cognitive load label in the training set to each cluster in the \ac{CLSE} dataset. These labels were used as a proxy to predict the cognitive load of utterances in the test and validation sets based on their cluster identity, effectively bypassing the use of any machine learning model for classification.

For the \ac{IEMOCAP} dataset, we analyzed the recording sessions based on the type of scenario (improvised or scripted). We computed the most probable label per session, obtained ASR transcripts with Whisper ASR, and incorporated the sentence embeddings using the LaBSE model. We then applied multi-class logistic regression for topic and emotion label classification, and repeated the experiments using ground truth transcriptions provided by the dataset authors.

\subsubsection{Evaluation}
We evaluated the impact of lexical overlap by comparing performance metrics using the Unweighted Average Recall (UAR). This metric offers a balanced assessment of performance across all classes, making it suitable for evaluating imbalanced datasets. We compared the performance of models trained on the original dataset splits with those trained on fixed (shuffled) splits, where lexical redundancy was minimized.

\subsection{CLSE}
\label{clse-experimental-setup}
We preprocessed, clustered and classified the \ac{CLSE} dataset as described in previous section. 

To further corroborate our hypothesis, we reshuffled the dataset splits based on cluster IDs. This process introduced speaker overlap but did not affect the cognitive load estimation, as each speaker experiences a full range of cognitive loads in the dataset.

\subsection{IEMOCAP}
As the structure of the \ac{IEMOCAP} database is different from \ac{CLSE}, we have adopted a slightly modified approach to verify our hypothesis. \ac{IEMOCAP} sessions are either improvised or scripted, with 12 different improvised scenarios and three scripts. We argue that the scenario of the recording session represents enough lexical context to the classifier that it can diminish the other factors (prosody etc.). 

To verify this claim, we conceived three different experiments: In the first one, we computed the most probable label per given session in a similar fashion to Section~\ref{clse-experimental-setup}. Then we obtained ASR transcripts with Whisper ASR (large-v3) and incorporated the sentence embeddings using the same extractor as in the \ac{CLSE} experiment. We then applied multi-class logistic regression for topic and emotion label classification.
Lastly, we repeated the same with ground truth transcriptions provided with the database.

All three experiments were conducted in five-fold cross-validation setup on a subset of the dataset comprising 5,502 sentences labeled as 'angry', 'happy', 'sad', and 'neutral'. The limited set of emotions is obtained as a standard pre-processing of \ac{IEMOCAP} and corresponds to the author's recommendations~\cite{IEMOCAP_reality_check}.

\section{Results}
\subsection{CLSE}
We assessed the influence of lexical overlap on classification performance using both original and fixed (shuffled) \ac{CLSE} data splits. The original \ac{CLSE} data displayed good performance metrics, but revealed patterns in the validation set that confirmed the lexical factors' role.

\begin{table}[ht]
\caption{Performance in Original vs Fixed CLSE Splits for Cluster-Based Classification} 
\label{tab:comparison_results_cluster}
\centering
\begin{tabular}{|l|c|c|}
\hline
& Original CLSE & Fixed CLSE \\
\hline
Train & 0.67 & 0.33 \\
\hline
Validation & 0.79 & 0.33 \\
\hline
Test & 0.57 & 0.33 \\
\hline
\end{tabular}
\end{table}

After minimizing lexical redundancy in shuffled splits, the performance metrics notably decreased to the chance level, as can be seen in Table \ref{tab:comparison_results_cluster}, further substantiating the lexical overlap's impact on classification.

We replicated the UBM-iVector system from the ComPaRE 2014~\cite{interspeech2014-compare} challenge, using a 64-component \ac{UBM} with a 50-dimensional iVector extractor and \ac{SVM} for classification~\cite{compare2014-ivector}. Using the same approach as in the cluster-based approach, we obtained results showing significant degradation of the performance on validation and test sets (Table~\ref{tab:UBM_results_comparison_UAR}).

\begin{table}[ht]
\caption{Performance in Original vs Fixed CLSE Splits for ComPaRE2014 System}
\label{tab:UBM_results_comparison_UAR}
\centering
\begin{tabular}{|l|c|c|}
\hline
Split & Original CLSE & Fixed CLSE \\
\hline
Train & 0.94 & 0.98 \\
\hline
Validation & 0.75 & 0.51 \\
\hline
Test & 0.64 & 0.51 \\
\hline
\end{tabular}
\end{table}

To corroborate our findings with a more recent speech-based system, we used HuBERT~\cite{hsu2021hubert} (version \textit{large-ll60k}) pre-trained on 60k hours from the Libri-Light dataset~\cite{libri-light} as a feature extractor. As a classifier, we used an \textit{Attentive correlation pooling} layer from~\cite{kakouros2022speechbased}. This classifier uses a multi-head attention layer, where attention weights resemble a mixture model with heads parametrizing the mixture component. The results of our experiments are shown in Table~\ref{tab:HuBERT_results_comparison_UAR}. They again manifest the same type of behavior as the previous experiments. 

\begin{table}[ht]
\caption{Performance in Original vs Fixed CLSE Splits for HuBERT Based System}
\label{tab:HuBERT_results_comparison_UAR}
\centering
\begin{tabular}{|l|c|c|}
\hline
Split & Original CLSE & Fixed CLSE \\
\hline
Train & 0.82 & 0.76 \\
\hline
Validation & 0.76 & 0.64 \\
\hline
Test & 0.74 & 0.52 \\
\hline
\end{tabular}
\end{table}

\subsection{IEMOCAP}
For IEMOCAP, the results of our experiments and comparable speech-based experiments taken from~\cite{10096808-IEMOCAP-Reality-Check} are shown in Table~\ref{tab:IEMOCAP_results}.

\begin{table}[ht]
\caption{Performance of Different Systems on IEMOCAP}
\label{tab:IEMOCAP_results}
\centering
\begin{tabular}{|l|c|}
\hline
System & UAR \\
\hline
\textit{Scenario based} & 0.62 \\
\textit{ASR transcriptions based} & 0.61 \\
\textit{Ground truth transcriptions based} & 0.61 \\
w2v2 based~\cite{emotion-recognition-wav2vec} & 0.67 \\
HuBERT based~\cite{yang2021superb} & 0.68 \\
MFCCs, Spectrogram, w2v2 based~\cite{zou2022speech-emotion-recognition} & 0.71 \\
\hline
\end{tabular}
\end{table}

These results highlight the robustness of textual features in emotion modeling but also invite further inquiry into the diminished role of paralinguistic elements. Our text-based experiments yielded only~$\sim9\%$ absolutely worse results than the state-of-the-art large pre-trained models, emphasizing the need for additional research to explore the interplay between textual and paralinguistic features in emotion recognition.

\section{Discussion and Conclusions}

This study reveals a critical, often-overlooked aspect of paralinguistic speech recognition: the significant lexical overlap in commonly used datasets. Our analysis of the CLSE and IEMOCAP datasets demonstrates that machine learning models may inadvertently learn text-dependent features rather than the targeted paralinguistic traits. This urges the community to reassess the integrity of current datasets and methodologies.

Reliance on ASR-focused pre-trained models like HuBERT risks conflating lexical and paralinguistic features. While excelling in text-dependent tasks, these models may obfuscate evaluations aimed at paralinguistic recognition.

While our results suggest that the lexical features in the speech data significantly influence the classification metrics, they do not negate the importance of paralinguistic features altogether. It is clear that when speech features are employed, indeed the performance of systems is better. They merely call for a more careful approach to evaluating paralinguistic recognition systems, with an explicit focus on decoupling textual and non-textual features.

We suggest that future work should focus on the development of methods for reducing text dependency in existing datasets. Additional evaluations should also be conducted using datasets that have been explicitly designed to minimize lexical overlap.

\bibliographystyle{IEEEtran}
\bibliography{IEEEabrv, refs.bib}

\begin{thebibliography}{10}
\providecommand{\url}[1]{#1}
\csname url@samestyle\endcsname
\providecommand{\newblock}{\relax}
\providecommand{\bibinfo}[2]{#2}
\providecommand{\BIBentrySTDinterwordspacing}{\spaceskip=0pt\relax}
\providecommand{\BIBentryALTinterwordstretchfactor}{4}
\providecommand{\BIBentryALTinterwordspacing}{\spaceskip=\fontdimen2\font plus
\BIBentryALTinterwordstretchfactor\fontdimen3\font minus
  \fontdimen4\font\relax}
\providecommand{\BIBforeignlanguage}[2]{{%
\expandafter\ifx\csname l@#1\endcsname\relax
\typeout{** WARNING: IEEEtran.bst: No hyphenation pattern has been}%
\typeout{** loaded for the language `#1'. Using the pattern for}%
\typeout{** the default language instead.}%
\else
\language=\csname l@#1\endcsname
\fi
#2}}
\providecommand{\BIBdecl}{\relax}
\BIBdecl

\bibitem{yap2012speech-egg-cognitive-load}
T.~F. Yap, ``Speech production under cognitive load: Effects and
  classification,'' Ph.D. dissertation, The University of New South Wales,
  2012.

\bibitem{Busso2008-iemocap}
\BIBentryALTinterwordspacing
C.~Busso, M.~Bulut, C.-C. Lee, A.~Kazemzadeh, E.~Mower, S.~Kim, J.~N. Chang,
  S.~Lee, and S.~S. Narayanan, ``{IEMOCAP}: interactive emotional dyadic motion
  capture database,'' \emph{Language Resources and Evaluation}, vol.~42, no.~4,
  pp. 335--359, Nov. 2008. [Online]. Available:
  \url{https://doi.org/10.1007/s10579-008-9076-6}
\BIBentrySTDinterwordspacing

\bibitem{compare2014-ivector}
M.~Van~Segbroeck, R.~Travadi, C.~Vaz, J.~Kim, M.~Black, A.~Potamianos, and
  S.~Narayanan, ``Classification of cognitive load from speech using an
  i-vector framework,'' Sep. 2014.

\bibitem{gallardo2019-saliency-based-lstm-on-clse}
A.~Gallardo-Antolín and J.~Montero, ``A saliency-based attention lstm model
  for cognitive load classification from speech,'' Sep. 2019, pp. 216--220.

\bibitem{wav2vec-clse}
P.~Hecker, A.~Kappattanavr, M.~Schmitt, S.~Moontaha, J.~Wagner, F.~Eyben,
  B.~Schuller, and B.~Arnrich, ``Quantifying cognitive load from voice using
  transformer-based models and a cross-dataset evaluation,'' Dec. 2022, pp.
  337--344.

\bibitem{Schuller2013-computational-paralinguistics}
\BIBentryALTinterwordspacing
B.~W. Schuller and A.~M. Batliner, \emph{Computational Paralinguistics}.\hskip
  1em plus 0.5em minus 0.4em\relax Wiley, Oct. 2013. [Online]. Available:
  \url{https://doi.org/10.1002/9781118706664}
\BIBentrySTDinterwordspacing

\bibitem{MaryZarate2015-linguistic-paralinguistic-features}
\BIBentryALTinterwordspacing
J.~M. Zarate, X.~Tian, K.~J.~P. Woods, and D.~Poeppel, ``Multiple levels of
  linguistic and paralinguistic features contribute to voice recognition,''
  \emph{Scientific Reports}, vol.~5, no.~1, Jun. 2015. [Online]. Available:
  \url{https://doi.org/10.1038/srep11475}
\BIBentrySTDinterwordspacing

\bibitem{pepino}
L.~Pepino, P.~Riera, L.~Ferrer, and A.~Gravano, ``Fusion approaches for emotion
  recognition from speech using acoustic and text-based features,'' in
  \emph{ICASSP 2020 - 2020 IEEE International Conference on Acoustics, Speech
  and Signal Processing (ICASSP)}, 2020, pp. 6484--6488.

\bibitem{emotions-corpora-study}
Y.~Li, Y.~Mohamied, P.~Bell, and C.~Lai, ``Exploration of a self-supervised
  speech model: A study on emotional corpora,'' in \emph{2022 IEEE Spoken
  Language Technology Workshop (SLT)}, 2023, pp. 868--875.

\bibitem{Daneman1980-rspan}
\BIBentryALTinterwordspacing
M.~Daneman and P.~A. Carpenter, ``Individual differences in working memory and
  reading,'' \emph{Journal of Verbal Learning and Verbal Behavior}, vol.~19,
  no.~4, pp. 450--466, Aug. 1980. [Online]. Available:
  \url{https://doi.org/10.1016/s0022-5371(80)90312-6}
\BIBentrySTDinterwordspacing

\bibitem{radford2022robust-whisper}
A.~Radford, J.~W. Kim, T.~Xu, G.~Brockman, C.~McLeavey, and I.~Sutskever,
  ``Robust speech recognition via large-scale weak supervision,'' 2022.

\bibitem{feng-etal-2022-language-labse}
\BIBentryALTinterwordspacing
F.~Feng, Y.~Yang, D.~Cer, N.~Arivazhagan, and W.~Wang, ``Language-agnostic
  {BERT} sentence embedding,'' in \emph{Proceedings of the 60th Annual Meeting
  of the Association for Computational Linguistics (Volume 1: Long
  Papers)}.\hskip 1em plus 0.5em minus 0.4em\relax Dublin, Ireland: Association
  for Computational Linguistics, May 2022, pp. 878--891. [Online]. Available:
  \url{https://aclanthology.org/2022.acl-long.62}
\BIBentrySTDinterwordspacing

\bibitem{IEMOCAP_reality_check}
\BIBentryALTinterwordspacing
N.~Antoniou, A.~Katsamanis, T.~Giannakopoulos, and S.~Narayanan, ``Designing
  and evaluating speech emotion recognition systems: A reality check case study
  with iemocap,'' in \emph{ICASSP 2023 - 2023 IEEE International Conference on
  Acoustics, Speech and Signal Processing (ICASSP)}.\hskip 1em plus 0.5em minus
  0.4em\relax IEEE, Jun. 2023. [Online]. Available:
  \url{http://dx.doi.org/10.1109/ICASSP49357.2023.10096808}
\BIBentrySTDinterwordspacing

\bibitem{interspeech2014-compare}
B.~Schuller, S.~Steidl, A.~Batliner, J.~Epps, F.~Eyben, F.~Ringeval, E.~Marchi,
  and Y.~Zhang, \emph{The INTERSPEECH 2014 Computational paralinguistics
  challenge: cognitive \& physical load}, Jan. 2014.

\bibitem{hsu2021hubert}
\BIBentryALTinterwordspacing
W.-N. Hsu, B.~Bolte, Y.-H.~H. Tsai, K.~Lakhotia, R.~Salakhutdinov, and
  A.~Mohamed, ``Hubert: Self-supervised speech representation learning by
  masked prediction of hidden units,'' 2021. [Online]. Available:
  \url{https://arxiv.org/abs/2104.03502}
\BIBentrySTDinterwordspacing

\bibitem{libri-light}
\BIBentryALTinterwordspacing
J.~Kahn, M.~Riviere, W.~Zheng, E.~Kharitonov, Q.~Xu, P.~Mazare, J.~Karadayi,
  V.~Liptchinsky, R.~Collobert, C.~Fuegen, T.~Likhomanenko, G.~Synnaeve,
  A.~Joulin, A.~Mohamed, and E.~Dupoux, ``Libri-light: A benchmark for asr with
  limited or no supervision,'' in \emph{ICASSP 2020}.\hskip 1em plus 0.5em
  minus 0.4em\relax IEEE, May 2020. [Online]. Available:
  \url{http://dx.doi.org/10.1109/ICASSP40776.2020.9052942}
\BIBentrySTDinterwordspacing

\bibitem{kakouros2022speechbased}
\BIBentryALTinterwordspacing
S.~Kakouros, T.~Stafylakis, L.~Mosner, and L.~Burget, ``Speech-based emotion
  recognition with self-supervised models using attentive channel-wise
  correlations and label smoothing,'' 2022. [Online]. Available:
  \url{https://arxiv.org/abs/2211.01756}
\BIBentrySTDinterwordspacing

\bibitem{10096808-IEMOCAP-Reality-Check}
N.~Antoniou, A.~Katsamanis, T.~Giannakopoulos, and S.~Narayanan, ``Designing
  and evaluating speech emotion recognition systems: A reality check case study
  with iemocap,'' in \emph{ICASSP 2023}, 2023, pp. 1--5.

\bibitem{emotion-recognition-wav2vec}
\BIBentryALTinterwordspacing
L.~Pepino, P.~Riera, and L.~Ferrer, ``Emotion recognition from speech using
  wav2vec 2.0 embeddings,'' \emph{CoRR}, vol. abs/2104.03502, 2021. [Online].
  Available: \url{https://arxiv.org/abs/2104.03502}
\BIBentrySTDinterwordspacing

\bibitem{yang2021superb}
S.-w. Yang, P.-H. Chi, Y.-S. Chuang, C.-I.~J. Lai, K.~Lakhotia, Y.~Y. Lin,
  A.~T. Liu, J.~Shi, X.~Chang, G.-T. Lin \emph{et~al.}, ``Superb: Speech
  processing universal performance benchmark,'' \emph{arXiv preprint
  arXiv:2105.01051}, 2021.

\bibitem{zou2022speech-emotion-recognition}
\BIBentryALTinterwordspacing
H.~Zou, Y.~Si, C.~Chen, D.~Rajan, and E.~S. Chng, ``Speech emotion recognition
  with co-attention based multi-level acoustic information,'' 2022. [Online].
  Available: \url{https://arxiv.org/abs/2203.15326}
\BIBentrySTDinterwordspacing

\end{thebibliography}
\begin{acronym}[UBM-GMM]
\acro{ML}{Machine Learning}
\acro{AI}{Artificial Intelligence}
\acro{ASR}{Automatic Speech Recognition}
\acro{MFCC}{Mel-Frequency Cepstral Coefficients}
\acro{UBM}{Universal Background Model}
\acro{GPU}{Graphics Processing Unit}
\acro{GID}{Gender Identification}
\acro{LID}{Language Identification}
\acro{SID}{Speaker Identification}
\acro{EGG}{Electroglottograph}
\acro{ANS}{Autonomic Nervous System}
\acro{SNS}{Sympathetic Nervous System}
\acro{PNS}{Parasympathetic Nervous System}
\acro{HPA}{Hypothalamic-Pituitary-Adrenal}
\acro{CLT}{Cognitive Load Theory}
\acro{ECG}{Electrocardiogram}
\acro{EDA}{Electro-Dermal Activity}
\acro{BP}{Blood pressure}
\acro{HR}{Heart rate}
\acro{SCR}{Skin Conductance Response}
\acro{GSR}{Galvanic Skin Response}
\acro{STAI}{State-Trait Anxiety Inventory}
\acro{STAI-Y2}{State-Trait Anxiety Inventory-Y2}
\acro{STAI-Y1}{State-Trait Anxiety Inventory-Y1}
\acro{PSS}{Perceived Stress Scale}
\acro{PSS14}{Perceived Stress Scale 14}
\acro{PSS10}{Perceived Stress Scale 10}
\acro{PSS4}{Perceived Stress Scale 4}
\acro{NASA-TLX}{NASA Task Load Experience}
\acro{WMC}{Working Memory Capacity}
\acro{OSPAN}{Operational Span task}
\acro{RSPAN}{Reading Span task}
\acro{TSST}{Trier Social Stress Test}
\acro{MAST}{Maastricht Acute Stress Test}
\acro{MAT}{Mental Arithmetic Task}
\acro{HIT}{Hand Immersion Task}
\acro{CPT}{Cold Pressor Test}
\acro{SUSC-0}{Speech under Stress Conditions-0}
\acro{SUSC-1}{Speech under Stress Conditions-1}
\acro{CoLoSS}{Cognitive Load Corpus with Speech and Performance Data from a Symbol-Digit Dual-Task}
\acro{CLSE}{Cognitive Load with Speech and EGG}
\acro{SUSAS}{Speech Under Simulated \& Actual Stress Database}
\acro{ATC}{Air Traffic Controller}
\acro{BESST}{Brno Extended Stress and Speech Test}
\acro{FIT-BUT}{Faculty of Information Technology - Brno University of Technology}
\acro{FIT}{Faculty of Information Technology}
\acro{AGC}{Automatic Gain Control}
\acro{UBM}{Universal Background Model}
\acro{LDC}{Linguistic Data Consortium}
\acro{DFT}{Discrete Fourier Transform}
\acro{DCT}{Discrete Cosine Transform}
\acro{PLDA}{Probabilistic Linear Discriminant Analysis}
\acro{JFA}{Joint Factor Analysis}
\acro{SRE}{Speaker Recognition}
\acro{UBM-GMM}{Universal Background Model - Gaussian Mixture Model}
\acro{HRV}{Heart Rate Variability}
\acro{NN}{Neural Network}
\acro{NLP}{Natural Language Processing}
\acro{NATO}{North Atlantic Treaty Organization}
\acro{PSYCH-PHIL-MUNI}{Department of Psychology at the Faculty of Arts at Masaryk University}
\acro{MUNI}{Masaryk University}
\acro{fps}{Frames per Second}
\acro{PCM}{Pulse Code Modulation}
\acro{BESSTiANNO}{BESST ANNOtation}
\acro{VEST}{Video Event Segmentation Tool}
\acro{CL}{Cognitive Load}
\acro{RECOLA}{REmote COLlaborative and Affective interactions}
\acro{RTC}{Real-Time Clock}
\acro{NTP}{Network Time Protocol}
\acro{DL}{Deep Learning}
\acro{MLT}{Mental Load Theory}
\acro{VAD}{Voice Activity Detection}
\acro{MBC}{Munich Biovoice Corpus}
\acro{IEMOCAP}{Interactive Emotional Dyadic Motion Capture}
\acro{LaBSE}{Language-Agnostic BERT Sentence Embeddings}
\acro{UAR}{Unweighted Average Recall}
\acro{AHC}{Agglomerative Hierarchical Clustering}
\acro{SVM}{Support Vector Machine}
\end{acronym}  

\end{document}